\documentclass[aps,pra,twocolumn,groupedaddress]{revtex4-1}
\usepackage{amsmath} 
\usepackage{graphicx}
\usepackage{dcolumn}
\usepackage{mathrsfs}
\usepackage{bm}
\usepackage{subfigure}
\usepackage{float}
\usepackage[colorlinks,linkcolor=blue,anchorcolor=blue,citecolor=blue,urlcolor=blue]{hyperref}

\begin{document}
\title{Continuous-variable pairwise entanglement based on optoelectromechanical system}
\author{Qi-Zhi Cai$^{1}$}
\author{Jin-Kun Liao$^{1}$}\thanks{E-mail: jkliao@uestc.edu.cn}
\author{Qiang Zhou$^{1,2,}$}\thanks{E-mail: zhouqiang@uestc.edu.cn}
\address{$^{1}$School of optoelectronic science and engineering, University of Electronic Science and Technology of China, Chengdu, Sichuan, China}
\address{$^{2}$Institute of Fundamental and Frontier Sciences, University of Electronic Science and Technology of China, Chengdu, Sichuan, China}

\date{\today }
\begin{abstract}
Inspired by the discrete-variable pairwise entanglement, in this work, we in theory analyze the continuous-variable pairwise entanglement between microwave modes based on a hybrid optoelectromechanical system, where the multi-pair microwave superconducting circuits simutaneously interact with each other via a mechanical resonator, which forms a Fabry-Perot cavity along with a standing mirror. With experimentally reachable parameter settings, wanted entanglement can be acheived when the pair number up to 10, and more is also available, which has the potential to be useful in quantum technologies where the demand for scalability and intergration is continuously increasing.
\end{abstract}

\pacs{Valid PACS appear here}
\maketitle

\section{Introduction}
Entanglement, a bizarre or even counterintuitive phenomenon predicted by quantum mechanics, has deeply influenced people's view on nature and widely stimulated innovations in technologies [1-11]. Generally, two different types of entanglement are used to explore the unknowns and innovative technologies, namely discrete-variable (DV) and continuous-variable (CV) entanglement. The discrete one entangles states from different qubit with two distinct states, for example, the polarization states of a photon and the energy states of quantized superconducting circuits. While the continuous one entangles continuous variables, like vibrational modes of resonators and quantized modes of the electromagnetic field. Due to the fact that the electromagnetic field plays a fundamental role in classical information science and technology, the CV regime quantum systems and the entanglement between them are relevant to various quantum technologies originated from classical world, such as communication, computing, simulation and sensing [12-18]. 

As the preparation and manipulation of unit or single quantum system developing and getting increasingly mature, the complex hybrid systems or networks with good scalability attract heated research interests [19-20]. In this case, various designs of structure or topology are emerged to entangle different units in hybrid systems, among them, pairwise regime is widely investigated, mainly due to its symmetrical characteristic for simulating quantum many-body systems, which is beneficial to develope quantum complex networks and quantum chemistry, and for enhancing the performance of other quantum technologies. However, the research of the pairwise entanglement primarily concentrates on DV regime [21-23], considering the close relationship between DV and CV entanglement, a question naturally arises: how does the pairwise entanglement behave in CV regime? Indeed, to analyse the pairwise entanglement between CV carrier, like the DV counterpart, suitable platforms are needed, owing to the successful theoretical prediction and experimental realization on entanglement between mechanical mode and electromagnetic fields with different frequency [24-36], optoelectromechanical system should be the one.

In this work, we study the pairwise entanglement between multi-pair microwave modes in CV regime based on an optoelectromechanical system. The considered hybrid system is a derivative one described in Ref.[37], where drum-head capacitor is optically coated to form one mirror of a Fabry-Pérot optical cavity and coupled with the optical mode by radiation pressure, at the same time, the vibrating drum-head membrane could capacitively interact with the microwave superconducting circuit, thus realizes the interface between light and microwave modes via a mechanical resonator. The extension in our model is that we consider the mechanical resonator not interact with one microwave superconducting circuit but with multi-pair microwave superconducting circuits. We show that, in such structure design, stationary and robust microwave-microwave pairwise entanglement can be generated with various microwave pair numbers and wide range microwave resonate frequencies. This work is expected to open the door for CV pairwise entanglement, especially based on optoelectromechanical platform, and to be useful in building complex hybrid quantum network by combining DV and CV pairwise design [38], due to the effective optoelectromechanical coupling with DV qubit, such as spin, atom, quantum dots and so on.

This paper is organized as follows. Sec. II shows the physical model of the system, with its quantum Langevin equations for describing the dynamics and their linearization. In Sec. III, we will derive the correlation matrix of the quantum fluctuations of the system in order to obtain the logarithmic negativity, which is considered as the entanglement measure in this work. We study the CV pairwise entanglement properties between microwave modes in Sec. IV, and Sec. V is for conclusion.

\section{Model}
\begin{figure}
	\centering
	\includegraphics[width=1\linewidth]{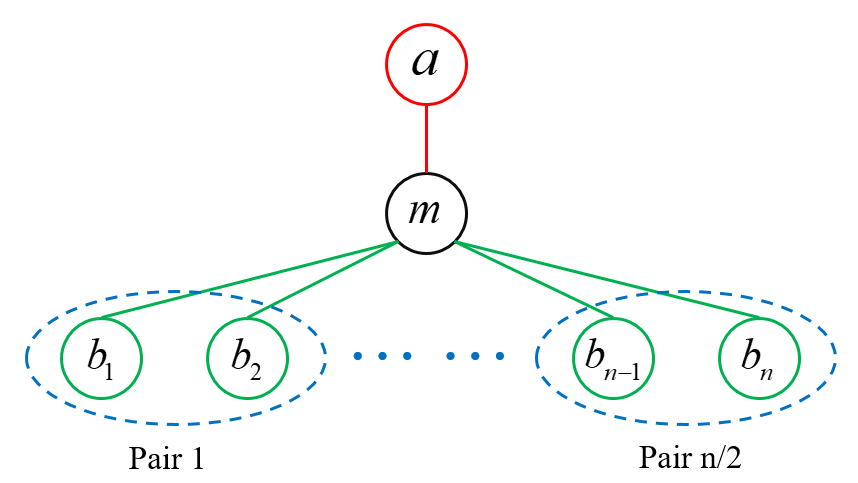}
	\caption{Simple picture for the hybrid system. $a$ is for the optical cavity, $m$ is for the mechanical resonator and $b_j$ are for the multi-pair microwave cavities.}
	\label{fig:Fig1.png}
\end{figure}
We consider a hybrid optotelectromechanical system containing $n$+2 units as shown in Fig.1, where $a$ represents Fabry-Pérot optical cavity (OC) with resonant frequency $\omega_c$, $m$ means mechanical resonator (MR) with resonant frequency $\omega_m$, and $b_j$ $(j=1,2,...,n,$ $n$ is even$)$ means the microwave cavity (MC) with resonant frequency $\omega_{b_j}$, in which two of the microwave modes can be regarded as a pair with suitable parameter settings, as the Results part will discuss; each MC is capacitively coupled with the MR simutaneously, at the same time, the MR is coupled with the OC via radiation pressure. The Hamiltonian of the system reads
\begin{equation}
\begin{split}
H =& \frac{{p_x^2}}{{2m}} + \frac{{m\omega _m^2{x^2}}}{2} + \hbar {\omega _c}{{a}^\dag } a - \hbar {G_{0c}}{{a}^\dag } a x\\
&+ \sum\nolimits_j {(\frac{{\Phi _j^2}}{{2{L_j}}} + \frac{{Q_j^2}}{{2[{C_j} + {C_{dj}}(x)]}} - {e_j}(t){Q_j}} )\\
&+ i\hbar {E_c}({{a}^\dag }{e^{ - i{\omega _{0c}}t}} - a{e^{i{\omega _{0c}}t}}),
\end{split}
\end{equation}
where $a$$(a^{\dagger})$ is the annihilation (creation) operator for OC, satisfying $[a,{a^\dag }]$=$1$, and $(x, p_x)$ are the canonical position and momentum of the MR with its effective mass $m$. $(\Phi_j, Q_j)$ are the canonical coordinates for the MCs, indicating the flux through effective inductors $L_j$ and the charge on effective capacitors $C_j$, respectively. ${G_{0c}} = ({{{\omega _c}} \mathord{\left/
		{\vphantom {{{\omega _c}} {L)}}} \right.
		\kern-\nulldelimiterspace} {l)}}\sqrt {{\hbar  \mathord{\left/
			{\vphantom {\hbar  {m{\omega _m}}}} \right.
			\kern-\nulldelimiterspace} {m{\omega _m}}}} $ is the coupling between OC and MR, with $l$ the length of the OC. The driving of the MCs is related to the electric potential ${e_j}(t) =  - i\sqrt {2\hbar {\omega _{wj}}{L_j}} {E_{wj}}({e^{i{\omega _{0wj}}t}} - {e^{ - i{\omega _{0wj}}t}})$, where ${E_{wj}} = \sqrt {{{2{P_{wj}}{\kappa _{wj}}} \mathord{\left/
			{\vphantom {{2{P_{wj}}{\kappa _{wj}}} {\hbar {\omega _{0wj}}}}} \right.
			\kern-\nulldelimiterspace} {\hbar {\omega _{0wj}}}}} $ with input microwave power $P_{wj}$ and damping rates $\kappa_{wj}$ for each MC. Similarly, ${E_c} = \sqrt {{{2{P_c}{\kappa _c}} \mathord{\left/
			{\vphantom {{2{P_c}{\kappa _c}} {\hbar {\omega _{0c}}}}} \right.
			\kern-\nulldelimiterspace} {\hbar {\omega _{0c}}}}} $ in which $P_c$ is the input laser power and $\kappa _c$ describes the damping rate of the OC. $C_{dj}(x)$ represents the capacitive interaction between MCs and MR, for convenience, we expand these functions around their equilibrium positions $d_j$, then ${C_{dj}}(x) = {C_{dj}}[1 - {{{x_j}(t)} \mathord{\left/
		{\vphantom {{{x_j}(t)} {{d_j}}}} \right.
		\kern-\nulldelimiterspace} {{d_j}}}]$. Expanding the microwave
	energy terms related to ${C_{dj}}(x)$ in Eq.(1) as Taylor series, we find the first order
\begin{equation}
\begin{split}
\frac{{Q_j^2}}{{2[{C_j} + {C_{dj}}({x_j})]}} = \frac{{Q_j^2}}{{2{C_{\Sigma j}}}} - \frac{{{\mu _j}}}{{2{d_j}{C_{\Sigma j}}}}{x_j}(t)Q_j^2,
\end{split}
\end{equation}
in which ${C_{\Sigma j}} = {C_j} + {C_{dj}}$ and ${\mu _j} = {{{C_{dj}}} \mathord{\left/
		{\vphantom {{{C_{dj}}} {{C_{\Sigma j}}}}} \right.
		\kern-\nulldelimiterspace} {{C_{\Sigma j}}}}$.
Under this condition, by introducing the annihilation and creation operator $(b,b^\dagger)$ of the MCs and the dimensionless position and momentum operators of the MR $(q,p)$, which satisfy $[{b_j},b_j^\dag ] = 1$ and $[q,p] = 1$, the Hamiltonian of Eq.(1) can be reshaped as follows
\begin{eqnarray}
\begin{split}
H =& \hbar {\omega _c}{a^\dag }a + \sum\limits_j {\hbar {\omega _{wj}}b_j^\dag {b_j}}  + \frac{{\hbar {\omega _m}}}{2}({p^2} + {q^2})\\
&- \sum\nolimits_j {\frac{{\hbar {G_{0wj}}}}{2}q{{({b_j} + b_j^\dag )}^2}}  - \hbar {G_{0c}}q{a^\dag }a\\
&- \sum\nolimits_j {i\hbar {E_{wj}}({e^{i{\omega _{0wj}}t}} - {e^{ - i{\omega _{0wj}}t}})({b_j} + b_j^\dag )} \\
&+ i\hbar {E_c}({a^\dag }{e^{ - i{\omega _{0c}}t}} - a{e^{i{\omega _{0c}}t}}),
\end{split}
\end{eqnarray}
where $\omega_{0c}$ and $\omega_{0wj}$ are driving frequencies of OC and MCs, repectively, and
\begin{equation}
\begin{split}
&{b_j} = \sqrt {\frac{{{\omega _{wj}}{L_j}}}{{2\hbar }}} {Q_j} + \frac{i}{{\sqrt {2\hbar {\omega _{wj}}{L_j}} }}{\Phi _j},\\
&q = \sqrt {\frac{{m{\omega _m}}}{\hbar }}x, p = \frac{{{{p}_x}}}{{\sqrt {\hbar m{\omega _m}} }},\\
&{G_{0wj}} = \frac{{{\mu _j}{\omega _{wj}}}}{{2{d_j}}}\sqrt {\frac{\hbar }{{m{\omega _m}}}}. 
\end{split}
\end{equation}
In the frame rotating at ${H_0} = \hbar {\omega _{0c}}{a^\dag } a + \sum\nolimits_j {\hbar {\omega _{0wj}} b_j^\dag {{b}_j}}$, the quantum Langevin equations (QLEs) describing the system dynamics read
\begin{equation}
\begin{split}
&\dot q = {\omega _m}p,\\
&\dot p =  - {\omega _m}q - {\kappa _m}p + {G_{0c}}{a^\dag }a + \sum\nolimits_j {{G_{0wj}}b_j^\dag {b_j}}  + \xi,\\
&\dot a =  - (i{\Delta _{0c}} + {\kappa _c})a + i{G_{0c}}qa + {E_c} + \sqrt {2{\kappa _c}} {a^{in}},\\
&{\dot b_j} =  - (i{\Delta _{0wj}} + {\kappa _{wj}}){b_j} + i{G_{0wj}}q{b_j} + {E_{wj}} + \sqrt {2{\kappa _{wj}}} {b^{in,j}},
\end{split}
\end{equation}
in which $\kappa_m$, $\kappa_a$ and $\kappa_{wj}$ are damping rates of MR, OC and MCs, respectively, and ${\Delta _{0c}} = {\omega _c} - {\omega _{0c}}$ and ${\Delta _{0w_{j}}} = {\omega _{w_{j}}} - {\omega _{0w_{j}}}$. $\xi(t)$ is the quantum Brownian noise acting on the MR, with the correlation function [39]
\begin{equation}
\begin{split}
{{\left\langle {\xi (t)\xi (t') + \xi (t')\xi (t)} \right\rangle } \mathord{\left/
		{\vphantom {{\left\langle {\xi (t)\xi (t') + \xi (t')\xi (t)} \right\rangle } 2}} \right.
		\kern-\nulldelimiterspace} 2} \approx {\kappa _m}(2{\bar n_m} + 1)\delta (t - t'),
\end{split}
\end{equation}
where we assume that $Q_m=\omega_{m}/\kappa_{m}\gg1$, which is valid in our model; $\bar{n}_m=1/[$exp$(\hbar\omega_m/k_B T)-1]$ is thermal excitation, $k_B$ is the Boltzmann constant and $T$ is the temperature of the MR. The optical and microwave input noises are given by $a^{in}$ and $b^{in,j}$,  respectively, which can be considered as zero-mean Gaussian, satisfying the correlation functions [40]
\begin{equation}
\begin{split}
&\left \langle a^{in}(t)a^{in,\dagger}(t')\right \rangle=[N(\omega_c)+1]\delta(t-t'),\\ 
&\left \langle a^{in,\dagger}(t)a^{in}(t')\right \rangle=N(\omega_c)\delta(t-t'),\\
&\left \langle b^{in,j}(t)b^{in,j,\dagger}(t')\right \rangle=[N(\omega_{wj})+1]\delta(t-t'),\\
&\left \langle b^{in,j,\dagger}(t)b^{in,j}(t')\right \rangle=N(\omega_{wj})\delta(t-t'), 
\end{split}
\end{equation}
where $N({\omega _c})=1/[$exp$(\hbar\omega_c/k_B T)-1]$ and $N({\omega _{wj}})=1/[$exp$(\hbar\omega_{wj}/k_B T)-1]$ are the mean thermal numbers of optical and microwaves fields, respectively. We can safely assume that $N(\omega_{wj})\simeq 0$ owing to ${{\hbar {\omega _c}} \mathord{\left/
		{\vphantom {{\hbar {\omega _c}} {{k_B}T}}} \right.
		\kern-\nulldelimiterspace} {{k_B}T}} \gg 1$, while $N({\omega _{wj}})$ cannot be ignored even the ambient temperature is quite low.

In our model, each MC and OC are intensely driven, which makes large amplitude for all cavities, i.e., $\left \langle O\right \rangle \gg 1$ $(O=a,b_j,q,p)$. Under this condition, we can linearize the dynamic around the semiclassical points of each cavity, by writing all operators as $O = O_s + \delta O$, where we have neglected high-order fluctuation terms. Then, we can focus on the linearized dynamics of system by inserting the approximation mentioned above into Eq.(5). By setting the derivatives to zero, the fixed semiclassical points for each subsystem read
\begin{equation}
\begin{split}
&{p_s} = 0,\\ 
&{q_s} = \frac{{{G_{0c}}{{\left| {{\alpha _s}} \right|}^2} + \sum\nolimits_j {{G_{0wj}}{{\left| {{\beta _{sj}}} \right|}^2}} }}{{{\omega _m}}},\\
&{\alpha _s} = \frac{{{E_c}}}{{{\kappa _c} + i{\Delta _c}}},\\
&{\beta _{js}} = \frac{{{E_{wj}}}}{{{\kappa _{wj}} + i{\Delta _{wj}}}}, 
\end{split}
\end{equation}
in which ${\Delta _c} = {\Delta _{0c}} - {G_{0c}}{q_s}$ and ${\Delta _{wj}} = {\Delta _{0wj}} - {G_{0wj}}{q_s}$ are the effective detuning of the optical and microwave fields, respectively, and linear QLEs for quantum fluctuations are given by
\begin{equation}
\begin{split}
&\delta \dot q = {\omega _m}\delta p,\\
&\delta \dot p =- {\omega _m}\delta q - {\kappa _m}\delta p + {G_{0c}}{\alpha _s}(\delta {a^\dag } + \delta a)\\
&+ \sum\nolimits_j {{G_{0wj}}{\beta _{sj}}(\delta b_j^\dag  + \delta {b_j})}  + \xi,\\
&\delta \dot a =  - ({\kappa _c} + i{\Delta _c})\delta a + i{G_{0c}}{\alpha _s}\delta q + \sqrt {2{\kappa _c}} {a^{in}},\\
&\delta {\dot b_j} =  - ({\kappa _{wj}} + i{\Delta _{wj}})\delta {b_j} + i{G_{0wj}}{\beta _{sj}}\delta q + \sqrt {2{\kappa _{wj}}} {b^{in,j}}, 
\end{split}
\end{equation}
where we have appropriately chosen phase references for all of the input optical and microwaves fields so that $\alpha_s$ and $\beta_{js}$ can be taken real and positive.

\section{Correlation matrix of the system and quantification of CV pairwise entanglement}
In order to obtaining the correlation matrix (CM) that calculates stationary CV entanglement between pairwise microwave modes, we shall introduce the quadratures $\delta X_c=(\delta a + \delta a^\dagger)/\sqrt{2}$ and $\delta Y_c=(\delta a - \delta a^\dagger)/i\sqrt{2}$ for OC, $\delta X_{wj}=(\delta b + \delta b^\dagger)/\sqrt{2}$ and $\delta Y_{wj}=(\delta b - \delta b^\dagger)/i\sqrt{2}$ for MCs, $\delta X^{c,in} =(\delta a^{in} + \delta a^{in,\dagger})/\sqrt{2}$, $\delta Y^{c,in} =(\delta a^{in} - \delta a^{in,\dagger})/i\sqrt{2}$, $\delta X^{wj,in} =(\delta b^{in,j} + \delta b^{in,j,\dagger})/\sqrt{2}$ and $\delta Y^{wj,in} =(\delta b^{in,j} - \delta b^{in,j,\dagger})/i\sqrt{2}$ for each corresponding noises, then the linear QLEs become
\begin{equation}
\begin{split}
&\delta \dot q = {\omega _m}\delta p,\\
&\delta \dot p =  - {\omega _m}\delta q - {\kappa _m}\delta p + {G_c}\delta {X_c} + \sum\nolimits_j {{G_{wj}}\delta {{\hat X}_{wj}}}  + \xi,\\
&\delta {\dot X_c} =  - {\kappa _c}\delta {X_c} + {\Delta _c}\delta {Y_c} + \sqrt {2{\kappa _c}} {X^{c,in}},\\
&\delta {\dot Y_c} = - {\Delta _c}\delta {X_c} - {\kappa _c}\delta {Y_c}  + {G_c}\delta q + \sqrt {2{\kappa _c}} {Y^{c,in}},\\
&\delta {\dot X_{wj}} =  - {\kappa _{wj}}\delta {X_{wj}} + {\Delta _{wj}}\delta {Y_{wj}} + \sqrt {2{\kappa _{wj}}} {X^{wj,in}},\\
&\delta {\dot Y_{wj}} =  - {\Delta _{wj}}\delta {X_{wj}} - {\kappa _{wj}}\delta {Y_{wj}} + {G_{wj}}\delta q + \sqrt {2{\kappa _{wj}}} {Y^{wj,in}}, 
\end{split}
\end{equation}
in which
\begin{equation}
\begin{split}
&{G_{wj}} = \sqrt 2 {G_{0wj}}{\beta _{sj}} = \frac{{{\mu _j}{\omega _{wj}}}}{{{d_j}}}\sqrt {\frac{{{P_{wj}}{\kappa _{wj}}}}{{m{\omega _m}{\omega _{0wj}}(\kappa _{wj}^2 + \Delta _{wj}^2)}}},\\
&{G_c} = \sqrt 2 {G_{0c}}{\alpha _s} = \frac{{2{\omega _c}}}{L}\sqrt {\frac{{{P_c}{\kappa _c}}}{{m{\omega _m}{\omega _{0c}}(\kappa _c^2 + \Delta _c^2)}}},
\end{split}
\end{equation}
are the effective the electromechanical and optomechanical couplings, respectively. Eq.(10) can rewrite in the form of
\begin{equation}
\begin{split}
\dot u(t) = Au(t) + n(t),
\end{split}
\end{equation}
where $u(t) = [\delta q,\delta p,\delta {{X}_c},\delta {{Y}_c},\delta {{X}_{w1}},\delta {{Y}_{w1}},...,\delta {{X}_{wn}},\delta {{Y}_{wn}}{]^T}$ (the notation $T$ means matrix transport) is a column vector with 2$n$+4 dimensions, the same as noise vector $n(t)$, and drift matrix $A$ is a (2$n$+4)$\times($2$n$+4) square matrix, whose explicit expression is shown in Appendix. As mentioned above, the dynamics describing the system are linear and the quantum noise terms in Eq. (12) are zero-mean Gaussian, so the steady state of the quantum fluctuations is a CV $n$+2 Gaussian state, completely characterized by a (2$n$+4)$\times($2$n$+4) CM, which can be attained by solving the Lyapunov equation [41-42]
\begin{equation}
\begin{split}
AV + V{A^T} =  - D,
\end{split}
\end{equation}
where D = Diag$[0,{\kappa _m}(2{\bar n_m} + 1),$${\kappa _c},{\kappa _c},{\kappa _{w1}}(2N({\omega _{w1}}) + 1),$${\kappa _{w1}}(2N({\omega _{w1}}) + 1),.,$${\kappa _{wn}}(2N({\omega _{wn}}) + 1),$${\kappa _{wn}}$$(2N({\omega _{wn}}) + 1)]$. As $V$ is obtained, the logarithmic negativity, i.e. the measure of CV entanglement in this work, of interested bipartite systems, like one pair of microwave modes, can be obtained by tracing out the rows and columns of other subsystems. With this operation, the reduced $4\times 4$ CM describing the interested bipartite pairwise subsystem is
\begin{equation}
\begin{split}
{V_{bi}} = \left( {\begin{array}{*{20}{c}}
	{{V_1}}&{{V_3}}\\
	{V_3^T}&{{V_2}}
	\end{array}} \right),
\end{split}
\end{equation}
and the entanglement between this pairwise modes is given by [43-44]
\begin{eqnarray}
E_{N}=max[0,-ln2\eta ^{-}],
\end{eqnarray}
where  ${\eta ^ - } \equiv {2^{{{ - 1} \mathord{\left/
				{\vphantom {{ - 1} 2}} \right.
				\kern-\nulldelimiterspace} 2}}}{[\Sigma ({V_{bi}}) - \sqrt {\Sigma {{({V_{bi}})}^2} - 4\det {V_{bi}}} ]^{{1 \mathord{\left/
				{\vphantom {1 2}} \right.
				\kern-\nulldelimiterspace} 2}}}$ and $\Sigma ({V_{bi}}) \equiv $det${V_1} + $det${V_2} - 2$det${V_3}$.

\section{Results}
\begin{figure}
	\centering
	\includegraphics[width=1\linewidth]{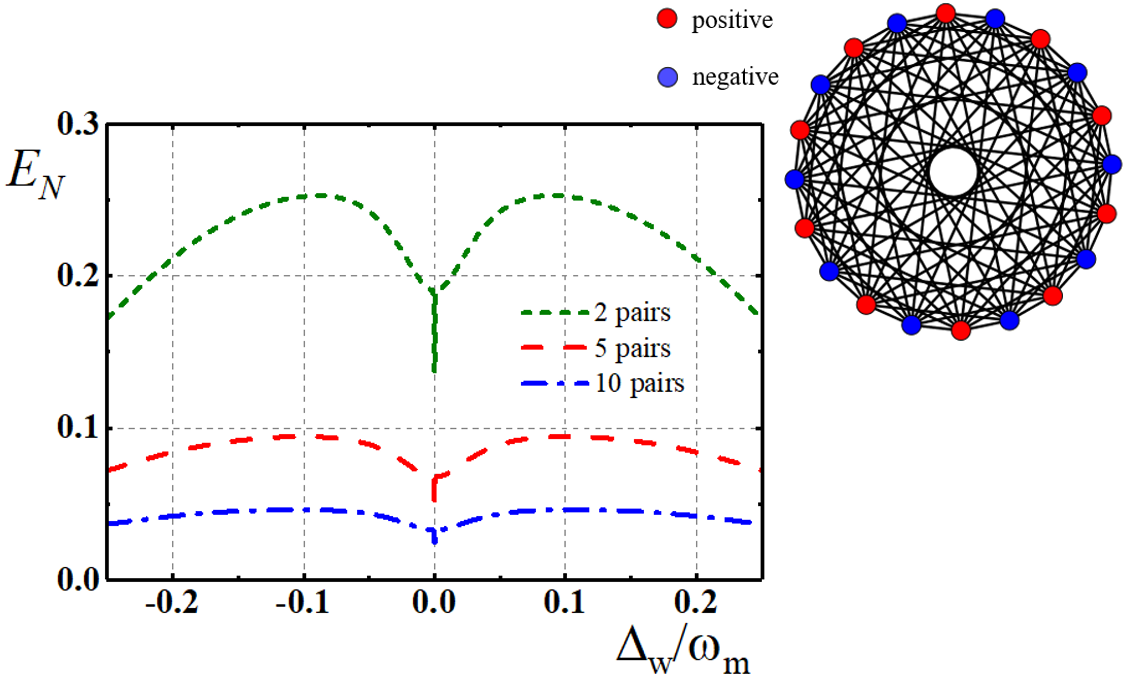}
	\caption{CV pairwise entanglement versus the uniform detuning $\Delta_w$ of MCs as the MC pair increases. Parameters: all MC resonant frequency is identical $\omega_c$=9 GHz, surrounding temperature T=15 mK. Green short dashed line shows the entanglement between two microwave modes with opposite detunings in the system with 2 pairs MCs, so as for red dashed line with 5 pairs MCs and blue dot dash line with 10 pairs MCs. The black line in the plot on the right shows the existence of entanglement between different MC in the system with 10 pairs MC. ``Positive" and ``negative" correspond to the MCs whose effective detuning with same or opposite sign compared to $\Delta_w$, respectively.}
	\label{fig:Fig2.png}
\end{figure}
\begin{figure}
	\centering
	\includegraphics[width=1\linewidth]{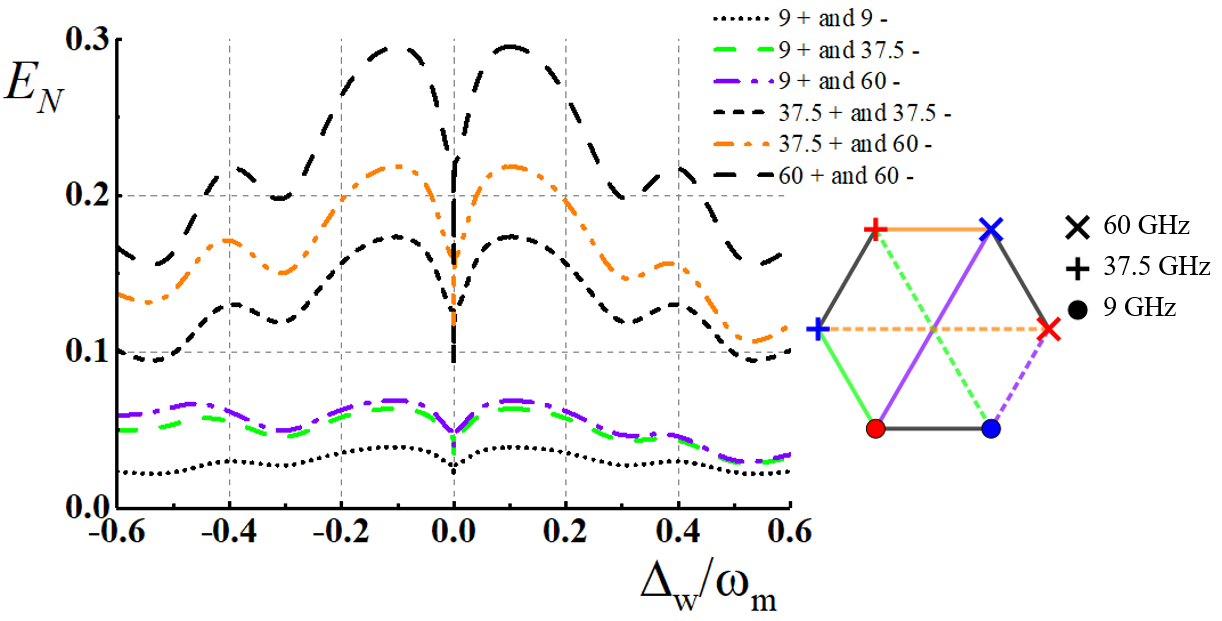}
	\caption{Pairwise entanglement between microwave modes with different resonant frequency in system containing 3 pair MCs, the first pair resonate at 9 GHz, 37.5 GHz for the second and 60 GHz for the third. 9+ or 9- means the MC with positive or negative detuning in 9 GHz microwave pair, the meaning of positive or negative is the same with Fig.2, so does for 37.5± and 60± and the surrounding temperature T=15 mK.}
	\label{fig:Fig3.png}
\end{figure}
In this section, we show the results of CV pairwise entanglement between microwave modes generated in a multipartite optoelectromechanical system. All results satisfy the Routh-Hurwitz criterion described in Appendix, which guarantes the steady state of the system. The parameter setting is a feasible extension over previous experimental works [37]: for OC, driving laser wavelegth $\lambda_{0c}$ = 1550 nm, damping rate $\kappa_c$ = 0.08 $\omega_m$, driving power $P_c$ = 30 mW and length of the cavity $L$ = 1 mm, optical effective detuning $\Delta_{c}=0.5\omega_m$, and for MR, effective mass $m$ = 10 ng, resonant frequency $\omega _m / 2 \pi$= 10 MHz and quality factor $Q$ = $5 \times {10^4}$. For MC part, as shown in Fig.1, we have several microwave pairs simutaneously interact with MR, the parameters of two MCs in each pair are almost same except effective detuning $\Delta_{wj}$ for satisfying the Routh-Hurwitz criterion, these two detunings are opposite, for instance, $\Delta_{w1}$ = $-\Delta_{w2}$, and every newly added microwave pairs should follow this criterion. Under this condition, the parameter regime between distinct microwave pairs can be different, which will not affect the stability of the system, in this work, for simplicity, we choose all microwave pairs with the same parameters except resonant frequency of each MC, and we will show its influence on the CV pairwise entanglement properties later. The parameters for MCs read: damping rates $\kappa_{w}$ = 0.02 $\omega_m$, input power $P_{w}$ = 30 mW, the parameter related to the electromechanical coupling $d$ = 100 nm and $\mu$ = 0.008, and the uniform effective microwave detunings $\Delta_{w}$ $ \equiv$ $\Delta_{w1}$ = $- \Delta_{w2}$ $=\cdot\cdot\cdot=$ $\Delta_{w(n-1)}$ = $- \Delta_{wn}$. Other unmentioned parameters are shown below each figure.

In Fig.2, we mainly focus on the influence of pair number on the pairwise entanglement of microwave modes, so we let the resonant frequency of all MC to be identical. In this case, all MC is classified into two categories with opposite detunings, and every MC in one category is exactly the same. The entanglement only exists between two MCs whose detunings are opposite as we will discuss later, in other word, choose one MC in each category arbitrarily, the entanglement will exist between these two MCs, and the entanglement will not exist between MCs picked from the same category as the right plot in Fig.2 shows. Back to pair number's influence, as the MC pair added, the pairwise entanglement declined, it can be phenomenologically thought that, as the scale of the system increases, the correlation between the subsystems deceases. Under this kind of structural design and parameter setting, we can further explore the maximum network size where CV entanglement still exists or useful, as long as the experimental conditions allow.
\begin{figure}[htp!]
	\centering
	\subfigure[]{
		\label{fig:Fig4(a).png}
		\includegraphics[width=1\linewidth]{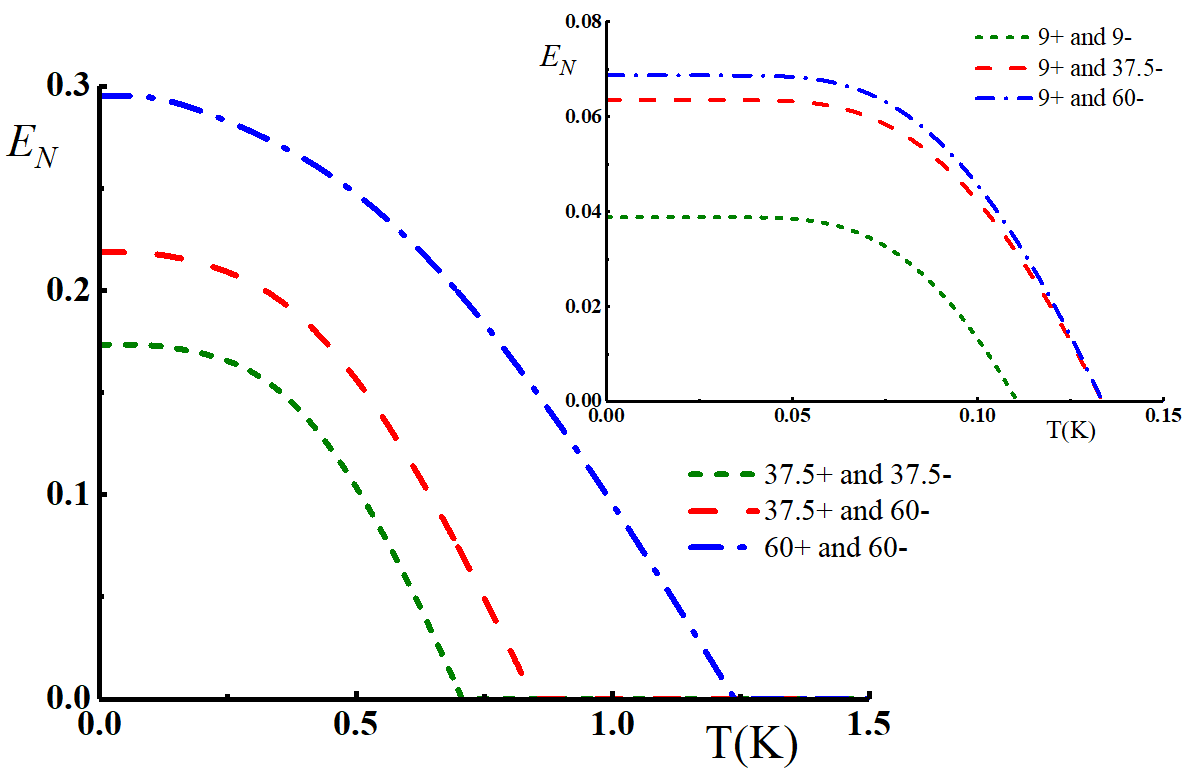}}
	\hspace{1in}
	\subfigure[]{
		\label{fig:Fig4(b).png}
		\includegraphics[width=0.8\linewidth]{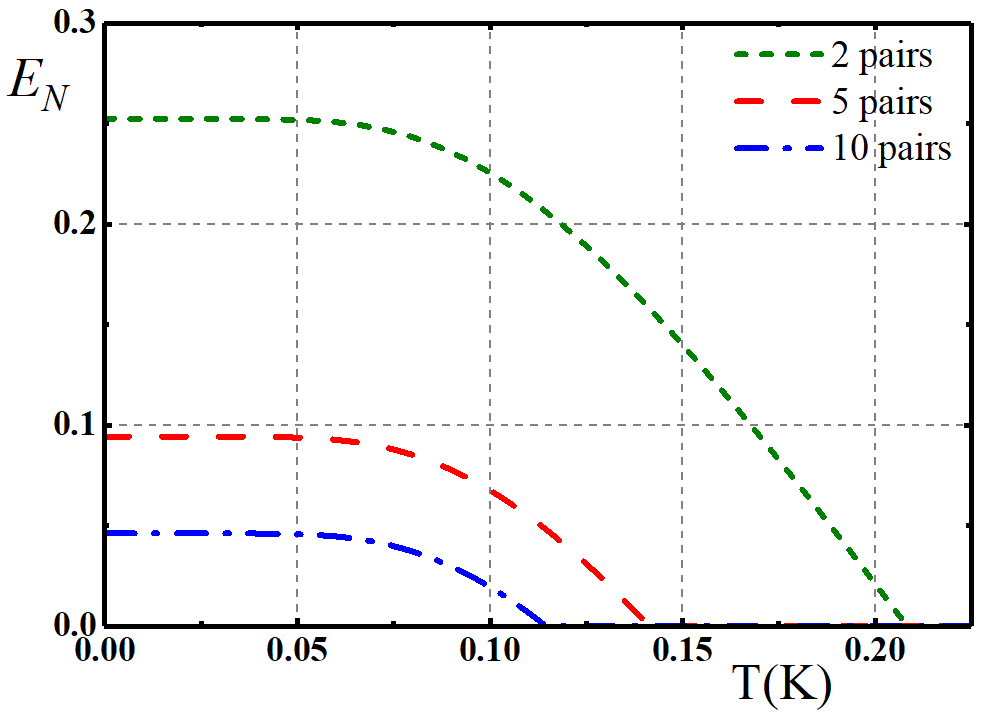}}
	\caption{Entanglement between two microwave modes versus temperature, (a) Model described in Fig.2, (b) Model described in Fig.3. $\Delta_w=-0.1\omega_m$.}
	\label{fig:Fig4.png}
\end{figure}

Then, we turn to study the CV pairwise entanglement when microwave pairs resonate at different frequency. For convenience, we choose the 3-pairs system, in which the first MC pair resonate at 9 GHz, 37.5 GHz for the second and 60 GHz for the third. Actually, modes with any frequency in microwave band can get entangled in our system, which is beneficial for fulfilling the requirements of broadband working in quantum communication and quantum computers. From Fig.3, we can know that the higher frequency of microwave modes, the larger entanglement between them can achieve, in part because higher frequency microwave photon contains more energy, which leads to stronger robustness in the thermal noise environment. We also notice that, because of structural symmetry, the entanglement between 9+ and 37.5- is the same as between 9- and 37.5+, so does for other pairs, thus we can entangle pairwise microwave modes with different frequency as long as their detunings are opposite, which is expected to be useful in distributed quantum sensing and quantum radar [16-17,42-43] working on the microwave band.
\begin{figure}
	\centering
	\includegraphics[width=1\linewidth]{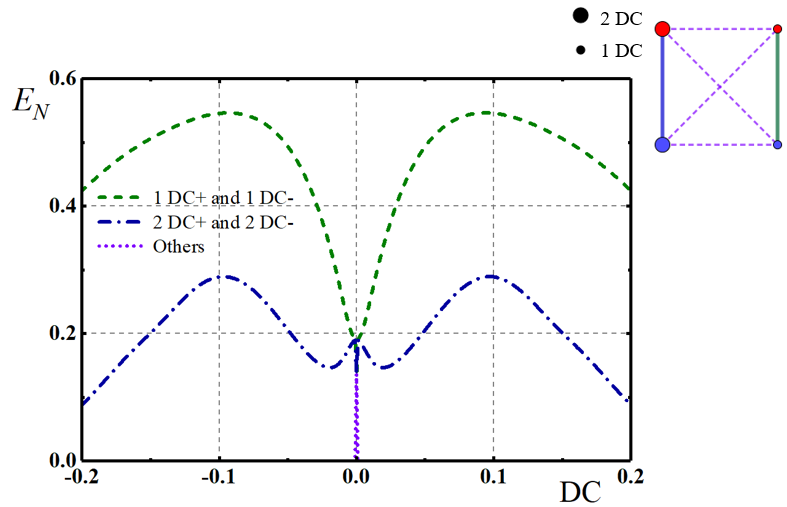}
	\caption{Entanglement between two microwave modes versus detuning coefficient (DC) in two-pair-MC system. The effective detuning of the first MC pair $\Delta_{w1}$ = $-\Delta_{w2}$ = DC$\cdot\omega_m$, of the second pair $\Delta_{w3}$ = -$\Delta_{w4}$ = 2DC$\cdot\omega_m$. The other parameter is the same with 2-pair model in Fig.2.}
	\label{fig:Fig5.png}
\end{figure}

Based on two models described above, we further investigate the relation between CV pairwise entanglement and temperature. As shown in Fig.4(a), the entanglement between higher-frequency microwave modes can represent stronger thermal robustness; in Fig.4(b), we found that the pairwise entanglement still survive above 100 mK in system containing 10 MC pairs, and the entanglement in more scalable system shows lower resistence in thermal environment. Both of them corresponds to the above statement.

Next, we study the influence of microwave effective detuning on pairwise entanglement. As shown in Fig.5, the effective detuning of the first pair MC $\Delta_{w1}$ = $-\Delta_{w2}$ = DC$\cdot\omega_m$, where the constant DC means the detuning coefficient, and the effective detuning for the second pair $\Delta_{w3}$ = $-\Delta_{w4}$ = 2DC$\cdot\omega_m$. We can find that the entanglement does not exist when the detunings of two MCs are not opposite, this is because the scattering phonons making two microwave modes correlated do not match to each other. Two microwave modes with this characteristic only get entangled when the detunings are zero, i.e. the are all in resonance, this is why the entanglement plots have a dip in this point as shown in Fig.2 and Fig.3. Using this feature, we can make a entanglement switch that turn on or off the entanglement between MCs by changing the detunings of each MC.

From the results shown above, we can summarize three discoverings in CV pairwise entanglement based on our model. At first, as the scability of the system increases, in other word more MC pair interact in the system, the pairwise entanglement between microwave modes decreases. Secondly, higher-frequency microwave modes can get better entangled and more resistant in thermal noise environment, which guides us to explore higher-frequency quantum electromechanics. The last one, the detunings of entangled two MCs should be opposite, which may assist us in building entanglement switches in system containing multi-pair MCs. We believe these findings will help us with designing CV pairwise system based on optoelectromechanical systems.

\section{Conclusion and outlook}
We have analyzed the CV pairwise entanglement with the help of well-studied optoelectromechanical system. Our model, from perspective in science, may have potential to investigate the quantum many-body system along with the DV systems, due to the strong coupling and effective state transfer between them. Furthermore, under the trend that the research platforms of quantum technologies are getting increasingly scalable and integrated, our scheme would play important role in various rapid-developing quantum technologies, with increasing scability and integration, like quantum-enhanced distributed sensing mentioned above and so on [18-19,45-46], which inspires us to realize this kind of optoelectromechanical system in lab to fulfill the huge demand in future.

\section{Appendix}
\begin{widetext}
The explicit expression of drift matrix $A$ is
\begin{equation}
\begin{split}
A = \left( {\begin{array}{*{20}{c}}
	0&{{\omega _m}}&0&0&0&0&0&0& \cdots &0&0&0&0\\
	{ - {\omega _m}}&{{\kappa _m}}&{{G_c}}&0&{{G_{w1}}}&0&{{G_{w2}}}&0& \cdots &{{G_{w(n - 1)}}}&0&{{G_{wn}}}&0\\
	0&0&{ - {\kappa _c}}&{{\Delta _c}}&0&0&0&0& \cdots &0&0&0&0\\
	{{G_c}}&0&{ - {\Delta _c}}&{ - {\kappa _c}}&0&0&0&0& \cdots &0&0&0&0\\
	0&0&0&0&{ - {\kappa _{w1}}}&{{\Delta _{w1}}}&0&0& \cdots &0&0&0&0\\
	{{G_{w1}}}&0&0&0&{ - {\Delta _{w1}}}&{ - {\kappa _{w1}}}&0&0& \cdots &0&0&0&0\\
	0&0&0&0&0&0&{ - {\kappa _{w2}}}&{{\Delta _{w2}}}& \cdots &0&0&0&0\\
	{{G_{w2}}}&0&0&0&0&0&{ - {\Delta _{w2}}}&{ - {\kappa _{w2}}}& \cdots &0&0&0&0\\
	\cdots & \cdots & \cdots & \cdots & \cdots & \cdots & \cdots & \cdots & \cdots & \cdots & \cdots & \cdots & \cdots \\
	0&0&0&0&0&0&0&0& \cdots &{ - {\kappa _{w(n - 1)}}}&{{\Delta _{w(n - 1)}}}&0&0\\
	{{G_{w(n - 1)}}}&0&0&0&0&0&0&0& \cdots &{ - {\Delta _{w(n - 1)}}}&{ - {\kappa _{w(n - 1)}}}&0&0\\
	0&0&0&0&0&0&0&0& \cdots &0&0&{ - {\kappa _{wn}}}&{{\Delta _{wn}}}\\
	{{G_{wn}}}&0&0&0&0&0&0&0& \cdots &0&0&{-{\Delta _{wn}}}&{ - {\kappa _{wn}}}
	\end{array}} \right).
\end{split}
\end{equation}
\end{widetext}
If the real part of all the eigenvalues of the drift matrix $A$ is negative, the system is stable and will approach a steady state. The exact calculation by Routh-Hurwitz theorem [42] is too cumbersome so we omit them here, and the solution of Eq.(12) is 
\begin{equation}
\begin{split}
u(t)=M(t)u(0)+\int_{0}^{t}dsM(s)n(s),
\end{split}
\end{equation}
where $M(t)$=exp$(At)$ and $n(t)$ is the noise vector.

The steady state of the system's quantum fluctuations are completely characterized by the $n\times n$ correlation matrix with its components ${V_{ij}} = {{\left\langle {{u_i}(\infty ){u_j}(\infty ) + {u_j}(\infty ){u_i}(\infty )} \right\rangle } \mathord{\left/
		{\vphantom {{\left\langle {{u_k}(\infty ){u_l}(\infty ) + {u_l}(\infty ){u_k}(\infty )} \right\rangle } 2}} \right.
		\kern-\nulldelimiterspace} 2}$. When the system is stable, the components become
\begin{equation}
\begin{split}
V = \int_0^\infty  {dsM(s)D{M^T}(s)},
\end{split}
\end{equation}
and $M(\infty ) = 0$. With the help of Lyapunov's first theorem [41], Eq.(17) is equivalent to Eq.(13): $AV + V{A^T} =  - D$.

\begin{acknowledgments}
This work has been supported by National Key R$\& $D Program of China (2018YFA0307400); National Natural Science Foundation of China (NSFC) (61775025, 91836102).
\end{acknowledgments}

\bibliography{zhoupra}

\end{document}